\documentstyle{article}

\normalsize

\def\e{\epsilon}
\def\f{\phi}
\def\g{\gamma}

\def\m{\mu}
\def\n{\nu}

\def\r{\rho}

\def\t{\tau}

\def\Hat#1{\rlap{\kern.10em$\widehat{\phantom G}$}#1}
\def\HAt#1{\rlap{\kern.05em$\widehat{\phantom G}$}#1}

\def\cAp#1{\rlap{\kern.1em$\widehat{\phantom{G\vrule height.8em}}$}#1{}}
\def\Cap#1{\rlap{\kern.05em$\widehat{\phantom{G\vrule height.8em}}$}#1{}}

\newcounter{sxn}

\newcounter{axn}

\def\br{\begin{thebibliography}{abc}}
\def\er{\end{thebibliography}}

\date{}

\begin{document}

\bibliographystyle{unsrt}
\footskip 1.0cm
\thispagestyle{empty}
\setcounter{page}{0}
\begin{flushright}
IC/94/18\\
Napoli: DSF-T-1/94, INFN-NA-IV-1/94\\
January 1994\\
\end{flushright}
\vspace{10mm}

\centerline {\LARGE VORTEX SOLUTIONS IN}
\vspace{5mm}
\centerline {\LARGE TWO-HIGGS-DOUBLET SYSTEMS }
\vspace*{15mm}
\centerline {\large G. Bimonte}
\vspace{5mm}
\centerline{ \small and }
\vspace*{5mm}
\centerline{ \large G. Lozano}
\vspace*{5mm}
\centerline {\it International Centre for Theoretical Physics, P.O.BOX 586}
\centerline {\it 34100 Trieste, ITALY}
\vspace*{25mm}
%\baselinestretch{2.0}
\normalsize
\centerline {\bf Abstract}
\vspace*{5mm}
{\large
We analyze the existence of string-like defects in a two-Higgs-doublet system
having $SU(2) \times U(1)_Y \times U(1)_{Y^{\prime}}$ as gauge group. We are
able to show that, when certain relations among the parameters hold, these
configurations satisfy a set of first order differential equations
(Bogomol'nyi equations) and their energy is proportional to their
topological charge.}
\newpage

\baselineskip=24pt
\setcounter{page}{1}
\newcommand{\be}{\begin{equation}}
\newcommand{\ee}{\end{equation}}

New interest in the study of string-like defects in spontaneously broken
gauge theories has arisen after the observation made by
Vachaspati \cite{vacha} that the embedding \cite{nambu}
of the Nielsen-Olesen \cite{niel}
string in the Standard Electroweak theory (model free of
topological defects) is stable for a certain range of parameters.
Nevertheless, the realistic values of the parameters, as derived
from experiments, lie outside this range of classical stability
\cite{james}.\\
This fact has led other authors to explore the existence of vortex solutions
in extended versions of the standard model, containing a richer Higgs
sector \cite{goran,la,peri}.
Dvali and Senjanovic have recently considered \cite{goran}
a two-Higgs-doublet model with an additional $U(1)$ global
symmetry which renders the vacuum manifold topologically non-trivial.
Due to the global character of this extra symmetry topological configurations
have (logarithmically) divergent energy.\\
In this letter, we will consider an $SU(2) \times U(1)_Y \times U(1)_{
Y^{\prime}}$ model, which corresponds to the model of \cite{goran},
once the additional $U(1)$ symmetry is gauged.\\
We will be interested in the stability of finite energy
strings in this model. The main result is that we are able to write a
topological bound for the energy (Bogomol'nyi bound \cite{bogo}) and, as a
consequence, the stability of the configurations satisfying the
bound is automatic. This bound is saturated when the fields satisfy
a set of first order differential equations (self-dual or
Bogomol'nyi equations, BE). As in the $U(1)$ case, the existence of BE
severely constrains the form of the potential. This would be and
unwanted feature of the model if it were not for the fact that in
all models studied so far the existence of BE signals the presence
of Supersymmetry (SUSY) \cite{witten}. In fact, our model has the same gauge
group structure as that of SUSY extensions of the Weinberg-Salam
Model that arise
as low energy limits of $E_6$ based grand unified or
superstring theories \cite{gun}.
 Nevertheless, the Higgs structure of these models is more
complicated than in our case, due to the presence of additional $SU(2)$
singlets. Here, we shall content ourselves with our model which is the
simplest extension of the Standard Model presenting BE and we leave for a
future publication \cite{bimo}
the analysis of the connection with realistic
SUSY extensions of the Standard Model.\\

The model we consider is described by the following Lagrangian density:
\be
{\cal L}= \frac{1}{4} W^a_{\mu\nu}W^{a \mu \nu}+
\frac{1}{4} B_{\mu\nu}B^{ \mu \nu}+\frac{1}{4} {\tilde B}_{\mu\nu}
{\tilde B}^{ \mu \nu}+\left| D^{(1)}_{\mu}\phi_1\right|^2
+\left| D^{(2)}_{\mu}\phi_2\right|^2 - V(\phi_1,\phi_2) \label{1}
\ee
where $W^a_{\mu\nu}, B_{\m\n}$ and ${\tilde B}_{\m\n}$ are the field
strengths associated with the gauge group
$SU(2) \times U(1)_Y \times U(1)_{Y^{\prime}}$ and the covariant derivatives
are defined as
$\footnote{repeated space-time and gauge-group indices are summed over,
while there is no summation on the index q, unless explicitly stated}$:
\be
D_{\m}^{(q)}\phi_q=\left( \partial_{\m} + \frac{i}{2}g \tau^a W^a_{\m}+
\frac{i}{2}g^{\prime} Y_q B_{\m}+\frac{i}{2}g_1 Y_q^{\prime}{\tilde B}_{\m}
\right) \phi_q .     \label{2}
\ee
In Eq.(2), $\t_a$ are the Pauli matrices,
 the index $q=1,2$ labels the Higgs doublets $\phi_q$ and $Y_q, Y^{\prime}_q$
denote, respectively, the $U(1)_Y$ and $U(1)_{Y^{\prime}}$ charges of
$\phi_q$.\\
The energy per unit length of axially symmetric static
configurations is then given by:
\be
E=\int d^2x \left[\frac{1}{4} W^a_{ij}W^a_{ ij}+
\frac{1}{4} B_{ij}B_{ij}+\frac{1}{4} {\tilde B}_{ij}
{\tilde B}_{ij}+\left| D^{(1)}_{i}\phi_1\right|^2
+\left| D^{(2)}_{i}\phi_2\right|^2 + V(\phi_1,\phi_2)\right] \label{3}
\ee
For the moment we will leave the potential $V(\phi_1,\phi_2)$
unspecified.\\
We start by using the standard Bogomol'nyi identity which consists in
writing
$$
\left| D^{(q)}_{i}\phi_q\right|^2=\frac{1}{2}\left| D^{(q)}_{i}\phi_q
-i \g_q \e_{ij}D^{(q)}_{j}\phi_q\right|^2 +
$$
\be
+\frac{1}{4}\g_q g \phi_q^{\dagger}
\tau^a \phi_q \e_{ij}W^a_{ij}+
\frac{1}{4}\g_q g^{\prime}Y_q
 \phi_q^{\dagger}
 \phi_q \e_{ij}B_{ij}+\frac{1}{4}\g_q g_1 Y^{\prime}_q \phi_q^{\dagger}
 \phi_q \e_{ij}{\tilde B}_{ij}+ \g_q\e_{ij} \partial_i J^{(q)}_j. \label{4}
\ee
where $\g_q^2=1$ and the current is defined by:
\be
J_j^{(q)}=\frac{1}{2i}\left[ \phi^{\dagger}D^{(q)}_j \phi_q-
(D^{(q)}_j \phi_q)^{\dagger}\phi_q \right]. \label{5}
\ee
After using (4) and assuming that the current goes to zero at infinity
, the energy can be rewritten as
$$
E=E_{ST}+\int d^2x \frac{1}{4}\left[W^a_{ij}+\e_{ij} R^a\right]^2+\frac{1}{4}
\left[B_{ij}+\e_{ij}R\right]^2+\frac{1}{4}
\left[{\tilde B}_{ij}+\e_{ij}{\tilde R}\right]
^2+
$$
$$
+ \frac{1}{2}\left| D^{(1)}_{i}\phi_1
-i \g_1 \e_{ij}D^{(1)}_{j}\phi_1\right|^2
+\frac{1}{2}\left| D^{(2)}_{i}\phi_2
-i \g_2 \e_{ij}D^{(2)}_{j}\phi_2\right|^2
+
$$
\be
+\left[V(\f_1,\f_2)-\frac{1}{2}R^a R^a
-\frac{1}{2}R^2-\frac{1}{2}{\tilde R}^2 \right]
   \label{6}
\ee
where
\be
R^a=\frac{g}{2}(\g_1 \phi^{\dagger}_1\tau^a \phi_1+
\g_2 \phi^{\dagger}_2\tau^a \phi_2)~~~,\label{7}
\ee
\be
R=\frac{g^{\prime}}{2}(\g_1 Y_1\phi^{\dagger}_1
 \phi_1+\g_2 Y_2\phi^{\dagger}_2 \phi_2-\rho)~~~,\label{8}
\ee
\be
{\tilde R}=\frac{g_1}{2}(\g_1 Y_1^{\prime}\phi^{\dagger}_1
 \phi_1+\g_2  Y_2^{\prime}\phi^{\dagger}_2 \phi_2 - {\tilde \r}) \label{9}
\ee
and
\be
E_{ST}=\frac{1}{4} \int d^2x \e_{ij}(g^{\prime}\r B_{ij}+
g_1 {\tilde \r}{\tilde B}_{ij}). \label{10}
\ee
Here, $\r$ and ${\tilde \r}$ are two arbitrary constants. We then see that
if we choose the potential such that:
$$
V(\f_1,\f_2)=\frac{1}{2}R^a R^a
+\frac{1}{2}R^2+\frac{1}{2}{\tilde R}^2=
$$
$$
=\frac{1}{8}(g^2+g^{\prime 2}Y_1^2+
{g_1}^2Y^{\prime 2}_1)(\phi_1^{\dagger}\phi_1)^2+
\frac{1}{8}(g^2+g^{\prime 2}Y_2^2+
{g_1}^2Y^{\prime 2}_2)(\phi_2^{\dagger}\phi_2)^2+
$$
$$
+\frac{1}{4}\g_1\g_2(g^{\prime 2}Y_1Y_2+{g_1}^2Y^{\prime}_1
Y^{\prime}_2-g^2)(\phi_1^{\dagger}\phi_1)(\phi_2^{\dagger}\phi_2)+
\frac{g^2}{2}\g_1\g_2(\phi_1^{\dagger}\phi_2)(\phi_2^{\dagger}\phi_1)+
$$
\be
-\frac{1}{4}\g_1(\r g^{\prime 2}Y_1+{\tilde \r}{g_1^2}Y^{\prime}_1)
(\phi_1^{\dagger}\phi_1)
-\frac{1}{4}\g_2(\r g^{\prime 2}Y_2+{\tilde \r}{g_1^2}Y^{\prime}_2)
(\phi_2^{\dagger}\phi_2)+\frac{\r^2 g^{\prime 2}}{8}+
\frac{{\tilde \r}^2 {g_1}^{2}}{8}~~, \label{11}
\ee
the energy becomes a sum of squares plus a boundary term and then:
\be
E \ge E_{ST}~~~.\label{12}
\ee
The bound is reached if and only if the following Bogomol'nyi
equations are satisfied:
\be
W^a_{ij}=-\e_{ij}R^a~~~, \label{13}
\ee
\be
B_{ij}=-\e_{ij}R~~~,  \label{14}
\ee
\be
{\tilde B}{ij}=-\e_{ij} {\tilde R}~~~, \label{15}
\ee
\be
D^{(1)}_i \phi_1=i \g_1 \e_{ij}D^{(1)}_j\phi_1~~~,   \label{16}
\ee
\be
D^{(2)}_i \phi_2=i \g_2 \e_{ij}D^{(2)}_j\phi_2~~~.   \label{17}
\ee
In order to have finite energy configurations, the strengths of the gauge
fields must vanish at infinity while the Higgs fields must satisfy the
conditions
\be
\lim_{r \rightarrow \infty}\f_1(r,\theta)=v_1(\theta)~~,
\lim_{r \rightarrow \infty}\f_2(r,\theta)=v_2(\theta)~~,\label{18}
\ee
where
\be
V[v_1(\theta),v_2(\theta)]=0~~.\label{19}
\ee
These conditions are  met by requiring that:
\be
R(v_1,v_2)=0~~~,  \label{20}
\ee
\be
{\tilde R}(v_1,v_2)=0~~~,  \label{21}
\ee
\be
R^a(v_1,v_2)=0~~~.  \label{22}
\ee
The first two equations fix $\rho$ and ${\tilde \rho}$ in terms of
$|v_1|^2$ and $|v_2|^2$:
\be
\r=(\g_1Y_1|v_1|^2+\g_2Y_2|v_2|^2)~~~,\label{23}
\ee
\be
{\tilde \r}=(\g_1Y^{\prime}_1|v_1|^2+\g_2Y^{\prime}_2|v_2|^2)~~~.\label{24}
\ee
On the other hand, Eq. (22) implies that
\be
|v_1|^2=|v_2|^2 \equiv \frac{v_0^2}{2} \label{25}
\ee
and also determines the relative direction of $v_1(\theta)$ and
$v_2(\theta)$.\\
In order to prove this, first notice that by means of a smooth gauge
transformation one can always set one of the Higgs fields constant (at
infinity):
\be
v_1=\frac{v_0}{\sqrt 2}\left(\matrix{ 0 \cr 1 \cr}\right)~~~~~~~
v_2=\frac{v_0}{\sqrt 2}\left(\matrix{ A(\theta) \cr B(\theta) \cr}\right)
{}~~. \label{26}
\ee
Eq. (22) then implies
$$
A^*B=0
$$
\be
1=\g_1 \g_2(|A|^2-|B|^2)~~~. \label{27}
\ee
There are two cases:
\be
{\rm case~1)}~~~\g_1\g_2=-1 ~~~~~~A=0~~~~~B=e^{i\chi(\theta)}~~,\label{28}
\ee
\be
{\rm case~2)}~~~\g_1\g_2=1 ~~~~~~A=e^{i\chi(\theta)}~~~~~B=0~~.\label{29}
\ee
Since case 2) can be obtained from case 1) via an operation of
charge conjugation on either one of the two Higgs fields , we will only
consider, in the sequel of the paper, case 1).\\
We now see that the $SU(2)$ scalar $w(\theta)
\equiv \frac{2}{v_0^2}
\f_1^{\dagger}\f_2$ defines a map
\be
w(\theta)=e^{i\chi(\theta)}:~~S_1 \rightarrow S_1~~.\label{30}
\ee
The winding $n$ of this map is the topological charge of the
configuration.\\
We will now show how the energy bound can be expressed in terms
of this winding number.
 We saw earlier that at
infinity the two Higgs fields have to be parallel. Via a smooth gauge
transformation it is always possible to put them in the form:
\be
\lim_{r \rightarrow \infty} \f_q=
\frac{v_0}{\sqrt 2}\left(\matrix{ 0 \cr e^{im_q \theta}
\cr}\right)~~~. \label{31} \ee
The topological charge is then equal to the relative winding of the
Higgs fields:
\be
n=m_2-m_1~~~.\label{32}
\ee
On the other hand since finite energy configurations are such that
\be
\lim_{r \rightarrow \infty} D^{(q)}\f_q=0    \label{33}
\ee
we obtain at infinity:
\be
2m_1-gW^3_{\theta}+g^{\prime}Y_1B_{\theta}+{g_1}Y^{\prime}_1
{\tilde B}_{\theta}=0~~~,\label{34}
\ee
\be
2m_2-gW^3_{\theta}+g^{\prime}Y_2B_{\theta}+{g_1}Y^{\prime}_2
{\tilde B}_{\theta}=0~~~.\label{35}
\ee
The energy bound
\be
E_{ST}=\frac{1}{2}\pi v_0^2\g_1[g^{\prime}(Y_1-Y_2)B_{\theta}+
{g_1}(Y^{\prime}_1-Y^{\prime}_2){\tilde B}_{\theta} ]  \label{36}
\ee
can then be expressed in terms of the topological charge as
\be
E_{ST}=\pi v_0^2\g_1(m_1-m_2)=-\pi v_0^2\g_1n~~.
\label{37}
\ee
Due to its topological character, $n-$vortex configurations saturating
the bound are then necessarily stable. Now, let us study in more
detail such configurations.
As we said, they satisfy the Bogomol'nyi Equations (13-17). The simplest
ansatz we can imagine is one where the Higgs are parallel in all space:
$$
W^1_i=W^2_i=0~~~~~~~W^3_r=B_r={\tilde B}_r=0~~~~,
$$
\be
\f_q=
\frac{v_0}{\sqrt 2}\left(\matrix{ 0 \cr f_q(r)e^{im_q \theta} \cr}\right)~~~,
\label{38}
\ee
where
$$
\lim_{r \rightarrow \infty}f_q=1~~.
$$
Eqs. (16-17) then become:
\be
\partial_r f_q^2=-\g_q \frac{1}{r}(2m_q-gW^3_{\theta}+
g^{\prime}Y_q B_{\theta}+g_1Y^{\prime}_q {\tilde B}
_{\theta})f_q^2~~~.\label{39}
\ee
On the other hand, by taking linear combinations of Eqs. (13-15), we obtain
\be
gg_1(Y_1Y^{\prime}_2-Y_2Y^{\prime}_1)W^3_{\theta}+gg_1(Y^{\prime}_2-
Y^{\prime}_1)B_{\theta}-g^{\prime}g(Y_2-Y_1){\tilde B}_{\theta}=0 \label{40}
\ee
Eqs. (39-40) allow us
to determine $B_{\theta},{\tilde B}_{\theta}$ and
$W^3_{\theta}$ once $f_1$ and $f_2$ are known. The equations for these
last fields are found by inserting (39) in (13-15),
\be
\nabla^2 \left(\matrix{ \log~f_1^2 \cr \log~f_2^2 \cr}\right)~=~M
\left(\matrix{f_1^2-1 \cr f_2^2-1 \cr}\right)~~~,\label{41}
\ee
where $M$ is the following mass matrix:
\be
M=\frac{v_0^2}{4}\left(\matrix{g^2+g^{\prime 2}Y_1^2+{g_1}^2Y^{\prime 2}_1
& -g^2-g^{\prime 2}Y_1 Y_2-{g_1}^2 Y^{\prime}_1 Y^{\prime}_2 \cr
-g^2-g^{\prime 2}Y_1 Y_2-{g_1}^2 Y^{\prime}_1 Y^{\prime}_2
&g^2+g^{\prime 2}Y_2^2+{ g_1}^2Y^{\prime 2}_2 \cr}\right)~~~.
\label{42}
\ee
Eq. (41) is the generalization, for the two doublets model, of
the $U(1)$ vortex equation \cite{bogo}:
\be
\nabla^2 \log f^2=m^2(f^2-1)~~~.\label{43}
\ee
Clearly, Eq. (41) only holds away of the zeroes of $f_1$ and $f_2$.
The behavior near the origin can be deduced from (39)
\be
f_q=c_qr^{-2\g_qm_q},~~~r \rightarrow 0 \label{44}
\ee
and it then follows that, in order to have regular solutions, one must
have
\be
\g_q m_q \le 0~~. \label{45}
\ee
As we are working with $\g_1\g_2=-1$, this implies
$$
m_1 m_2\le 0~~~.
$$
As for the behavior at infinity, we can write
\be
f_q(r)=1+h_q(r)~~~~,~~~|h_q| \ll 1 \label{46}
\ee
and then
\be
\nabla^2 \left(\matrix{h_1 \cr h_2\cr}\right)~=~M
\left(\matrix{h_1 \cr h_2 \cr}\right)~~~.\label{47}
\ee
This implies that:
$$
f_q = 1+c_q K_0(m_- r)~~~,  \label{48}
$$
where $m_-$ is the lowest eigenvalue of the mass matrix.\\
Notice that within our ansatz there is a degeneracy
associated with the different possible splitting of the topological
charge between the two Higgs. In fact, for any $n$  there are $|n|+1$
different assignments of $m_1$ and $m_2$ compatible with Eq. (45). It is
easy to verify that solutions with different $(m_1,m_2)$, although
topologically
equivalent are not related by a gauge transformation.\\
To conclude, let us analyze the mass spectrum of the theory. This can be
easily done in the unitary gauge
\be
\f_1=\frac{1}{\sqrt 2}\left(\matrix{ \chi \cr v_0+h^0_1 \cr}\right)~~~~~~~
\f_2=\frac{1}{\sqrt 2}\left(\matrix{ -\chi \cr v_0+h^0_2 \cr}\right)~~.
\label{49} \ee
where $\chi$ is a complex field and $h^0_1$ and $h^0_2$ are real. With the
potential given by (11), one finds:
\be
m^2_{\chi}=\frac{1}{2}g^2v_0^2     \label{50}
\ee
while the masses of the neutral components can be obtained by
diagonalizing the mass matrix $M$.
Regarding the gauge boson sector, one can check that besides the photon
there is one charged particle,$W^+$, and two neutral ones, $Z_1$ and
$Z_2$, whose masses are two by two equal to those of the corresponding
Higgs particles. This phenomenon is the analogue of the one occurring
in the $U(1)$ case where Bogomol'nyi equations exist only when the mass
of the Higgs is equal to the mass of the gauge boson.\\
Summarizing, we have been able to show the existence of
stable string-like solutions of arbitrary topological charge. The
result follows from the possibility of writing a topological bound
for the energy. We have proposed a simple ansatz for the solutions
of the BE, and even in this case they do not correspond to an
embedding of the Nielsen-Olesen vortex. It would be interesting to
know if there are more general solutions, exhibiting in particular
the phenomenon of $W$-condensation \cite{amb}. Finally, the most interesting
open problem is to understand the connection of our model with
realistic SUSY extensions of the Standard Model. This issue is under
current investigation \cite{bimo}.\\

{\bf Acknowledgments}
\vspace{\medskipamount}

We would like to thank G.Senjanovic for interesting discussions
and Prof. Abdus Salam, the International Atomic Energy Agency and
UNESCO for hospitality at the International Centre
for Theoretical Physics.

\end{document}